# A Theory of the Knowledge Industry

Hisham Ghassib
ghassib@psut.edu.jo
The Princess Sumaya University for Technology (PSUT),
P.O.Box 1438, Al-Jubaiha 11941, Jordan

**Abstract**
This paper deals with the social production of knowledge in the exact sciences. After defining the term, exact science, it delineates the broad dynamic of its history. It, then, offers a socio-economic historical explanation of why the production of knowledge has become a major industry, if not the largest industry, in the last hundred years. The paper concludes by drawing a detailed blueprint of the components, mechanisms and specificities of the knowledge industry.

Keywords: Knowledge production, Knowledge industry, Exact science,, Scientific revolution, Antique science, Epistemic heritage.

**Introduction**
Knowledge is an amazing human power. It is almost a divine attribute of man, a divine flash of light, a sacred divine fire, stolen by Prometheus to grace man with it. It is of course a real power-- nay, the basis of all human powers. It is multifaceted and basically abstract, which makes its study a particularly complicated, hard and multifaceted enterprise. Since discursive knowledge is a distinctive human attribute, which distinguishes man from other living organisms, we know that it must be related to the complexity of the human brain and its interactive abilities with nature and society. But, how? Where does it emanate from? How is it born, and how does it persist? Our starting point in this work is the historical observation that knowledge is not meta-human, that it is not pre-existent, and, thus, that it is not merely a process of recollection or inspiration.

Rather, it is socially produced. There is no innate, eternal knowledge pre-existing in the human mind or elsewhere, as in Plato and to a certain extent in Descartes, or in some divinely written eternal text, as in religion, or in the human heart, as in mystics. Knowledge is neither recollected nor read from a primary source, nor yet mystically intuited. Rather, human societies tend to produce it socially on the bases of their productive powers and evolving needs (Ghassib, 1992). Thus, the genesis, existence, development and evolution of knowledge must be approached socio-historically. Admittedly, knowledge is rooted in the structure of the human brain, but it is rooted in it as a possibility, not as a direct actuality. In actuality, it is a socio-historical product, which starts and develops historically, and in relationship with the totality of human history. Social man produces knowledge as he produces his life and as he survives (Ghassib, 1993;2011) This production of knowledge progresses as human society develops. Thus, knowledge has a history, within the context of the totality of human history.

In this study, we wish to elaborate a theory of knowledge production dealing with the following points. As we intend to focus on exact science, we shall start by giving an operationally useful definition of exact science. Next, we shall give an account of the broad structure of the history of exact science. In particular, we shall divide the history of natural science into two broad eras: antique natural science and modern natural science. These two eras are separated by a transitional revolutionary era, which is commonly called the scientific revolution of the 17th century. This categorization will lead us to the concept of the knowledge industry. We shall offer a socio-economic historical explanation of the transformation of natural science from a minor marginal enterprise into a major industry. We shall then embark on a detailed explication of the components, mechanisms and specificities of the knowledge industry.

**Exact Science**

When we use the term, exact science, we usually have in mind modern natural sciences-- in particular, physics, chemistry, astronomy, and geology. The essence of exact science is the dialectical relationship between mathematized theory and precise measurement, observation and experimentation. By a dialectical relationship, we mean an existential, necessary, transformational and contradictory relationship. Thus, exact science entails a specific way of theorization, which is mathematized in its very essence, and which can be concretized into a specific practice oriented towards precise measurement.

**From Antique to Modern Science**
Was exact science, in the sense outlined above, a modern European invention? Unfortunately, this erroneous view was the prevalent view in the West for a long time. It held that exact science was marginally born in ancient Greece, and later erupted in its totality in modern Europe. Modern scholarship in the history of science has shaken this erroneous idea to its core (Teresi, 2002). It has shown that the first truly exact science started in Babylon around the year 500 BC (Aaboe, 2001). That was Babylonian astronomy, which was characterized by a dialectical relationship between a very original form of numerical analysis and precise astronomical measurement. The next great instance of exact science was Greek astronomy, which started in the 2nd century BC by Hipparchus. Unlike Babylonian astronomy, it employed geometrical, rather than numerical analytic, methods. Hipparchus was of course influenced by Babylonian astronomy (Neugebauer, 1964), and his astronomy flourished in Egypt and the Near East, rather than in the West. It reached its Greek climax in Ptolemy in Alexandria in the 2nd century AD (Barbour, 2001). This climax was revived in the 8th and 9th centuries AD. It was soon to be absorbed, updated and corrected by Arabic astronomers. In the tenth century AD, a critical movement of Ptolemaic astronomy started in both the Arab East and the Arab West (Saliba, 1994; Saliba, 2007; Sabra, 1998). In subsequent centuries, Arabic astronomers attempted to construct alternative , but equally accurate, models, to the Ptolemaic models, more in accordance with Aristotelian principles. This movement lasted till the 16th century and beyond. Thus, Ptolemaic astronomy, which was a veritable exact science, reached its zenith in Arabic science. Another exact science started in Arabic civilization in the tenth century AD-- namely, Arabic optics (Rashed, 2003; Colic, 2007). The Arabic philosopher, Al-Kindi, Ibn Al-Sahl and Al-Hazen (Al-Hassan bin Al-Haitham) succeeded in turning the study of light into an exact science, whereby mathematical theorization was wedded to quantitative experimentation. One of its major fruits was the precise mathematical formulation of the sine law of refraction by Ibn Al-Sahl (Rashed, 2003). This law was later wrongly attributed to Snell and Descartes. However, prior to the modern era, exact science failed to become universal. That is, it failed to penetrate major areas of nature, such as the motion of material bodies, the constitution of matter and cosmology. The latter remained chained to unscientific, inexact, metaphysical methods. The prevailing methodology of natural knowledge was Aristotelian logico-metaphysical methodology, which was quintessentially non-mathematical and non-experimental. It seems that the mathematical-cum-observational method was confined to pure ideal entities, such as light and celestial objects. Matter was considered too gross and ponderous to be treated mathematically. The essence of the Aristotelian logico-metaphysical method was emphasizing the reality of first metaphysical principles on the one hand, and the senses on the other, whilst establishing a syllogistic bridge between these two ontological realms (Shea, 1972). Other non-scientific methods were also employed in pre-modern cultures, such as Pythagoreanism, Gnosticism, Mysticism, Hermeticism, Magic, Alchemy and Astrology. Thus, in pre-modern cultures, exact science was marginal and limited to a number of pockets dispersed into a sea of non-scientific methodologies. They failed to expand at the expense of their surrounding. This ultimately led to their contraction and the liquidation of the whole knowledge production enterprise in Arabic culture in particular. In pre-modern cultures, the production of knowledge was characterized by the following general traits:

1- It was a small enterprise compared to other sectors, such as the agricultural, commercial, handicraft, religious and military sectors, in terms of the size of investment and the number of its practitioners. The number of professional knowledge producers was minuscule compared to the huge number of peasants, traders, handicraftsmen and soldiers.

2- Its impact on the economy was almost non-existent, compared to the other economic sectors.

3- It had a guild-like organization.

4- As stated previously, it was limited in extent and marginal in effect.

5- It lacked self-autonomy, in that it was subjected to external-- religious and political-- authority in its very content. Its ultimate reference point was not scientific reason, but revelation.

In the 17th century, a momentous intellectual revolution occurred, which turned everything upside down, and broke all the confining limits of antique science (Kuhn, 1959; Koestler, 1959; Koyre, 1973; Margolis, 2002). In particular, Galileo succeeded in extending the Archimedean mathematical method, which had been so successful in astronomy and optics, to the motion of bodies (Machamer, 1998). He invented a new method-- some sort of a Platonic empirical method-- which enabled him to mathematize nature (Drake, 1991). At the same time, Kepler succeeded in inventing a new (admittedly, a crude) field physics to celestial phenomena, which enabled him to discover the correct laws of planetary motion (Stephenson, 1987). Following these two giants of the scientific revolution, Descartes laid down a materialist ontology for modern science (Descartes, 1969). Newton was to unify all that into a grand mathematical engine for producing and organizing natural knowledge (Bernard Cohen, 1980) These revolutionary events generated modern natural science. It was basically a comprehensive critical campaign, which volcanically extended the limited pockets of exact science at the expense of the metaphysical problematic that had prevailed. Whereas in pre-modern cultures, the pockets co-existed with the metaphysical bulk, in modern culture, the pockets exploded and destroyed the metaphysical bulk. It led to the invention of a new science (Bernard Cohen, 1968). which came to replace antique science entirely. The consequences were truly enormous. The new science led to a new technology and a new mode of production. It was seminal in transforming the nature and tempo of technology and material production. Without the new science, both the technological and the industrial revolutions would not have been possible. The scientific revolution ultimately transformed knowledge production from a marginal craft into a major, if not the major, industry in the modern world. Thus, compared to antique science, modern science has acquired the following general traits:

1- It has become a central enterprise.
2- It has become an industry, and probably the largest industry to boot. It is certainly one of the largest industries, if not the largest, in terms of the amount of investment and the number of employees in it, not to mention the quality of its employees.
3- It has become vital for the modern economy. That is, it is not a mere industry amongst equally important others, but has acquired the status of a strategic industry, just like oil and the arms industry.
4- It has receded from its previous guild-like form and become industrially structured.
5- It has acquired a noticeable degree of self-autonomy. It is no longer affiliated to any religious or political authority. It does not recognize any external authority. The only authority it recognizes is internal to itself-- namely, scientific reasoning or scientific reason.
6- In pre-modern cultures, science was guided by metaphysics and theology. In modern culture, the opposite relationship prevails; philosophy and theology are guided by science.

In short, for millennia, science was either non-existent or marginal. In the last two or three centuries, it has been transformed into a central concern, into one of the largest handful of industries. This is very peculiar. Why has this gigantic transformation occurred? What has happened in the last few centuries to nullify the lull that lasted for millennia? How can we explain this great upheaval?

**Towards A Political Economy Of Knowledge Production**
We attribute the relatively recent emergence of the knowledge industry, or of knowledge as an industry, to certain distinguishing features of the modern age, and the modern economy in particular.

These distinguishing features, which truly distinguish the modern economy from all past forms and modes of economic production, have demanded allotting knowledge a central place in modern society. We particularly identify the following two new features of modernity, which offer a satisfactory explanation of the emergence of the knowledge industry.

First, an essential feature of the capitalist mode of production is that it cannot survive and persist if it does not constantly revolutionize the implements of labour-- that is, the means of production, exchange, distribution, communication, transport, defense and aggression. In other words, a continually changing and rapidly developing technology is necessary for the reproduction of the modern economy. This was not the case in pre-modern economies, which tended to be conservative, and to uphold tradition at the expense of change.

Thus, pre-modern technology, which was based on intuitive, practical life knowledge-- the knowledge of the carpenter, sailor, blacksmith, hunter and peasant-- tended to be static, and accorded perfectly with the conservative essence of pre-modern economies. However, capitalism has demanded right from the start an altogether different technology, a rapidly developing technology in a state of permanent revolution (Marx, 1973). Such a technology cannot be based on practical life acquaintance. It can only be based on a universal extended science-- that is, on the ever expanding knowledge of causal principles and their effects. Such a technology can only be a practical embodiment of a knowledge industry, a central scientific enterprise. This means that the persistence of the modern economy and its expanded reproduction require a huge knowledge industry.

Second, unlike pre-capitalist formations, the modern economy is not geared towards the satisfaction of fixed needs, but towards the generation of new needs. Since it is profit oriented, it feeds and reproduces itself via need generation. The sustainability of profit making requires need generation. The leading need generator in the modern economy is the knowledge industry. Thus, the very essence of the modern economy requires the scientific enterprise. The latter is the motive force of the former. In short, modern capitalism created the knowledge industry, because the latter is necessary for its survival and reproduction.

**The Components and Mechanisms of the Knowledge Industry**
Being an industry, the knowledge industry can be analyzed and understood in terms of the general form and structure of industries. Like other industries, it consists of sites of production, producers, instruments of production, raw materials, methods of production, products and objects of production. Let us concretely analyze each of these components and locate the specificities of the knowledge industry within this context.

*1- Sites of Knowledge Production, or Knowledge Factories:*
The main sites of knowledge production are universities, research institutes and research units within companies and enterprises. The last hundred years have witnessed an exponential growth of these knowledge factories in terms of their number and sizes. At the beginning of the 20th century, research institutes were almost non-existent (Kragh, 2002), and universities were very limited in number and size. Yet, by the end of that century, the number of universities and research institutions has risen to tens of thousands, and their sizes have grown many times over, with a concomitant growth in budgets. The growth has been stunning indeed.

## 2- *The Knowledge Producers:*

The knowledge producers are principally a network of highly trained minds called scientists. They are supported by an army of administrators, engineers and technicians, but they are the direct producers. They are trained to deal with the raw material of knowledge, work with it and on it, and transform it into new knowledge. Knowledge is produced by teams of scientists within the context of a scientific community or institutional network. Thus, scientists are trained to deal with the raw material of knowledge, as well as to communicate with each other and with society at large. In this case, individual creativity is rooted in collective creativity, and conditioned by it.

## 3- *The Instruments of Knowledge Production:*

There are two broad classes of instruments of knowledge production: material implements and intellectual implements. Under the former, we list labs, mechanical and electronic devices, computers, calculators, softwares and suchlike. Under the latter, we list mathematical techniques, logical methods, and philosophical ideas. Scientists use such instruments creatively to work on, and with, the raw material of knowledge production.

## 4- *The Raw Material of Knowledge Production:*

The raw material of knowledge production is knowledge itself. To be more precise, it is the epistemic heritage of the scientific community. The epistemic heritage is a historically conditioned body of concepts, ideas, experimental, mathematical and logical techniques, theories, hypotheses, conjectures, philosophical ideas, experimental and observational facts and data, problematics, and practices recognized by the scientific community as its field of action and thought. At no time can the epistemic heritage be considered a logical whole, a wholly logically coherent structure a la Euclidean geometry, for instance. It aspires to become a theory of everything (Weinberg, 1993), or a wholly coherent logical edifice. This quest animates its progress. However, in actuality, it is a logically and conceptually inhomogeneous body of thought and practices, ridden with contradictions, absences, defects, uncertainties, problems, ungrounded conjectures and incoherencies. The scientist works with this complex body and on it, and interacts with the object of knowledge with it, in order, to solve its problems, resolve its contradictions, fill in its absences, remove its defects, increase its body of evidence, ground its conjectures, confirm its hypotheses, establish deeper connections and unify its seemingly unrelated elements. This is the basic mechanism, whereby the epistemic heritage develops, progresses and approaches the ideal of logical coherence.

## 5- *The Object of Knowledge:*

The epistemic heritage is not a closed universe, a closed whole. It is not some sort of a universal Hegelian Reason animated and evolving under the pressure of its internal contradictions (Hegel, 1977). Rather, it is more of an open totality constantly interacting with the object of knowledge via scientific practice, scientists' practices. Thus, the epistemic heritage is necessarily intentional. It is an open curve, that is closed, not in itself, but by the object of knowledge. It is necessarily constantly oriented towards the object of knowledge. It is concerned with such an object. Knowledge is ultimately produced by the systematic interaction of a scientific community with the object of knowledge. The former does not interact with the latter directly, but, rather, via and with the epistemic heritage. This means that the epistemic heritage does not develop merely under the pressure of its constant internal contradictions, but, rather, under the pressure of its varying internal and external contradictions. Its interaction with its object of knowledge feeds, as it were, its internal contradictions, moves them forward, and develops them further. The motive forces that animate the epistemic heritage are scientific practice in itself and its systematic interaction with the object of knowledge. This is indeed the dialectic of scientific development (Ghassib, 1988). In the case of physics, the object of knowledge is nature, or the universe. In the case of the social sciences, it is human society. The relationship between the epistemic heritage and the object of knowledge constitutes its axis of significance and meaning.

## 6- *The End Product of Knowledge Production:*

The very special specificity of the production of knowledge resides precisely in the peculiarity of its end product. The latter differs from other industrial products in two unique characteristics. First, the process of evaluation of the end product is an integral part of the production process itself. The end product is a mere potentiality until it is subjected to evaluation by the scientific community. The potential can only turn into the actual via a strict process of evaluation. The latter is a necessary component of the value of an end product of knowledge production. Without it, the value of the end product is not realized. If the evaluation is negative, the end product is considered worthless-- that is, it is not incorporated into the epistemic heritage, and is not considered a real end product of scientific practice-- even if the knowledge producers have spent years on its production. That explains why promotion criteria tend to emphasize scientific works published in internationally recognized journals, and why scientists aspire to publish their work in such journals.

Second, the end product of knowledge production is necessarily and intrinsically unenvisageable a priori. In conventional industries, the end product is envisageable beforehand. We know beforehand the type of end product we are aiming at. In fact, the method and process of production in this case is determined by this prior vision. The latter is translated beforehand into a set of pre-determined steps. However, in the case of the knowledge industry, this cannot possibly be done. A new end product cannot possibly be known a priori. Otherwise, it would fail to constitute a new end product. If the knowledge end product is to be genuinely new and original, then it cannot possibly be envisageable beforehand. It may sometimes be anticipated, but it cannot be fixed beforehand. This is the only type of industrial product that bears this peculiar characteristic.

*7- The Methods of Knowledge Production:*

From the sixth component, it follows that, strictly speaking, there are no fixed prior methods of production in the scientific enterprise. We cannot translate scientific practice into a set of pre-ordained, pre-determined, assembly-line-like steps. Scientific production is never purely routine work. It is intrinsically creative. It cannot be other than creative. Strictly speaking, there is no scientific method. This is a myth concocted by some philosophers. There is an infinite variety of scientific methods. Every major scientist creates his own method. Every major discovery and theoretical breakthrough requires a methodological innovation. Methodology is our key to unlock the secrets of nature. However, every secret requires its own key. A specific method may be suitable for a particular discovery, but it may not be so for another. In a sense, methodology is the other side of theory. It is the active side of theory. Every method has its limit. Once this limit is reached, transcending the prevailing method becomes a necessity. In science, methods are constantly being modified, overthrown and invented. Of course, scientific methods are inter-related. After all, scientists learn their methods from each other. However, as they progress in their problem-solving work, they are impelled to modify or transcend the method they have learnt. Thus, scientific methods evolve. Some sort of a natural selection process is at work here. The so-called scientific method is an organism that evolves.

## Conclusion and Prospects

In this paper, we have introduced the concept of the knowledge industry, suggested a mechanism for its genesis and emergence,

grounded this mechanism in the general march of history, particularly in socio-economic processes, and elaborated a productivist model of knowledge detailing the main components and mechanisms of the knowledge industry. Two fundamental questions arise from this elaboration. First, if there is an infinite variety of scientific methods, can we demarcate scientific practice from other social practices, including artistic, philosophical, religious, magical, astrological, alchemical, and mystical practices? If so, how? In a subsequent paper, I will show how this can be done by exploring the concept of scientific reason and the epistemological, ontological, ethical and socio-educational grounds of scientific practice. The second question concerns the concept of creativity in science. If science is fundamentally a creative activity, how do we characterize this creativity? In a future publication, I will address this question, and propose a theory grounding individual creativity in what I call collective creativity.